\documentclass[12pt]{article}
\usepackage{amsfonts}
\usepackage{amssymb}
\usepackage{amsmath}
\usepackage[latin1]{inputenc}

\newtheorem{theorem}{Theorem}

\newtheorem{lemma}[theorem]{Lemma}

\newtheorem{remark}[theorem]{Remark}

\newenvironment{proof}[1][Proof]{\textbf{#1:} }{\ \rule{0.5em}{0.5em}}
\begin{document}
\title{ Two-sided estimates for stock price distribution densities in jump-diffusion models}
\author{Archil Gulisashvili $\cdot$ \\ Josep Vives}
\date{}
\maketitle
\noindent \textbf{Abstract}\,\, We consider uncorrelated Stein-Stein, Heston, and Hull-White models and their perturbations by compound Poisson processes 
with jump amplitudes distributed according to a double exponential law. Similar perturbations of the Black-Scholes model
were studied by S. Kou. For perturbed stochastic volatility models, we obtain two-sided estimates for the stock price 
distribution density and compare the tail behavior of this density before and after perturbation. It is shown that if the value of the parameter, 
characterizing the right tail of the double exponential law, is small, then the stock price density in 
the perturbed model decays slower than the density in the original model. On the other hand, if the value of this parameter is large, 
then there are no significant changes in the behavior of the stock price distribution density.
-------------------------------- \\
\footnotesize The research of the second author was supported by grant MTM2009-07203 \\
------------------------------------- \\
\footnotesize Archil Gulisashvili \\
Department of Mathematics, Ohio University, Athens, OH 45701, USA \\
e-mail: guli@math.ohiou.edu \\
\\
Josep Vives \\
Departament de Probabilitat, L\`{o}gica i Estad\'{\i}stica, Universitat de Barcelona,
Gran Via 585, 08007-Barcelona (Catalunya), Spain \\
e-mail: josep.vives@ub.edu \\
\normalsize
\section{Introduction} 
It is assumed in the celebrated Black-Scholes model that the volatility of a stock is constant. However,
empirical studies do not support this assumption. In more recent models, the volatility of a stock is represented by a stochastic process. 
Well-known examples of stochastic volatility models are the Hull-White, the Stein-Stein, and the Heston model. The volatility processes in these models
are a geometric Brownian motion, the absolute value 
of an Ornstein-Uhlenbeck process, and 
a Cox-Ingersoll-Ross process, respectively. For more information on stochastic volatility models, see \cite{FPS} and \cite{G06}.
 
A stock price model with stochastic volatility is called uncorrelated if standard Brownian motions driving the stock price equation 
and the volatility equation are independent. 
In \cite{GS06}, \cite{GS09-2}, and \cite{GS09-3}, sharp asymptotic formulas were found for the distribution density of the stock price in uncorrelated Hull-White, Stein-Stein, and Heston models. Various applications of these formulas were given in 
\cite{GS09-1} and \cite{G}. The results obtained in \cite{GS09-2} and \cite{GS09-3} will be used in the present paper. 

It is known that the stock price distribution density in an uncorrelated stochastic volatility model possesses a certain structural symmetry (see formula 
(\ref{E:sym}) below). This implies a similar symmetry in the Black-Scholes implied volatility, which does not explain the volatility skew observed in practice. To improve the performance of an uncorrelated model, one can either assume that the stock price process and the volatility process are correlated, or add a jump component to the stock price equation or to the volatility equation. The stock price distribution in the resulting model fits the empirical stock price distribution better than in the uncorrelated case. However, passing to a correlated model or adding a jump component may sometimes lead to similar effects or may have different consequences (see e.g. \cite{ALV07} and \cite{ALPV08}).
Examples of stock price models with jumps can be found in 
\cite{B96}, \cite{K2}, and \cite{KW04}. We refer the reader to \cite{CT04} for more information about stock price models with jumps. An interesting discussion of the effect of adding jumps to the Heston model in contained in \cite{KR08}.

An important jump-diffusion model was introduced and studied by Kou (see \cite{K2} and \cite{KW04}). This model can be described as a perturbation of the Black-Scholes model by
a compound Poisson process with double-exponential law for the jump amplitudes. In the present paper, we consider
similar perturbations of stochastic volatility models. Our main goal is to determine whether significant changes may occur in the tail behavior of the stock price distribution after such a perturbation. We show that the answer depends on the relations between the parameters defining the original model and the characteristics of the jump process.
For instance, no significant changes occur in the behavior of the distribution density of the stock price in a perturbed Heston or Stein-Stein model 
if the value of the parameter characterizing the right tail of the double exponential law is large. On the other hand, if this value is small, then 
the distribution density of the stock price in the perturbed model decreases slower than in the original model. For the Hull-White model, 
there are no significant changes in the tail behavior of the stock price density, since this density decays extremely slowly.

We will next briefly overview the structure of the present paper. In Section \ref{FP}, we describe classical stochastic volatility models and their perturbations by a compound Poisson process. In Section \ref{MR} we formulate the main results of the paper and discuss what follows from them. Finally, in Section \ref{P}, we prove the theorems formulated in Section \ref{MR}.

\section{Preliminaries}\label{FP}
In the present paper, we consider perturbations of uncorrelated Stein-Stein, Heston, and Hull-White models by compound Poisson processes. 
Our goal is to determine whether the behavior of the stock price distribution density in the original models changes after such a perturbation. 

The stock price process $X$ and the volatility process $Y$ in the Stein-Stein model 
satisfy the following system of stochastic differential equations:
\begin{equation}
\left\{\begin{array}{ll}
dX_t=\mu X_tdt+\left|Y_t\right|X_tdW_t\\
dY_t=q\left(m-Y_t\right)dt+\sigma dZ_t.
\end{array}
\right.
\label{E:SS}
\end{equation}
This model was introduced and studied in \cite{SS91}. The process $Y$, solving the second equation in (\ref{E:SS}), is 
called an Ornstein-Uhlenbeck process. We assume that $\mu \in {\mathbb R}$, $q\ge 0$, $m\ge 0$, and $\sigma> 0$. 

The Heston model was developed in \cite{H93}. In this model, the processes $X$ and $Y$ satisfy 
\begin{equation}
\left\{\begin{array}{ll}
dX_t=\mu X_tdt+\sqrt{Y_t}X_tdW_t\\
dY_t=q\left(m-Y_t\right)dt+c\sqrt{Y_t}dZ_t,
\end{array}
\right.
\label{E:H}
\end{equation}
where $\mu\in {\mathbb R},$ $q>0$, $m\ge 0$, and $c>0$. The volatility equation in (\ref{E:H}) is uniquely solvable in the strong sense, 
and the solution $Y$ is a non-negative stochastic process. This process is called a Cox-Ingersoll-Ross process.

The stock price process $X$ and the volatility process $Y$ in the Hull-White  model 
are determined from the following system of stochastic differential equations:
\begin{equation}
\left\{\begin{array}{l}
dX_t=\mu X_tdt+Y_tX_t dW_t\\
dY_t=\nu Y_tdt+\xi Y_tdZ_t.
\end{array}
\right.
\label{E:sde1}
\end{equation}
In (\ref{E:sde1}), $\mu\in {\mathbb R},$ $\nu\in\mathbb{R}$, and $\xi> 0$. 
The Hull-White model was introduced in \cite{HW87}. The volatility process in this model is a geometric Brownian motion.

It will be assumed throughout the paper that standard Brownian motions $W$ and $Z$ in (\ref{E:SS}), 
(\ref{E:H}), and (\ref{E:sde1}) are independent. The initial conditions for the processes 
$X$ and $Y$ will be denoted by $x_0$ and $y_0$, respectively.

We will next discuss perturbations of the models defined above by a compound 
Poisson process with jump amplitudes distributed according to a double exponential law. 
Perturbations of the Black-Scholes model by such 
jump processes were 
studied by Kou in \cite{K2} and by Kou and Wang in \cite{KW04}. 
Some of the methods developed in \cite{K2} will be used in the present paper.   

Let $N$ be a standard Poisson process with intensity $\lambda>0$, and consider a compound Poisson process defined by 
\begin{equation}
J_t=\sum_{i=1}^{N_t} (V_i-1),\quad t\ge 0,
\label{Poi}
\end{equation}
where $V_i$ are positive independent identically distributed random variables that are 
independent of $\left\{N_t\right\}_{t\ge 0}$. It is also assumed that the distribution density $f$ of 
$U_i=\log V_i$ is double exponential, that is, 
\begin{equation}
f(u)=p\eta_1 e^{-\eta_1 u}{1\!\!1}_{\{u\geq 0\}}+q\eta_2 e^{\eta_2 u}{1\!\!1}_{\{u<0\}}.
\label{DE}
\end{equation}
where $\eta_1>1,$ $\eta_2>0,$ and $p$ and $q$ are positive numbers such that $p+q=1.$

Consider the following jump-diffusion stochastic volatility models:
\begin{equation}
\left\{\begin{array}{ll}
d{\widetilde X}_t=\mu {\widetilde X}_{t-}dt+\left|Y_t\right|{\widetilde X}_{t-}dW_t+{\widetilde X}_{t-} dJ_t\\
dY_t=q\left(m-Y_t\right)dt+\sigma dZ_t
\end{array}
\right.
\label{E:SS1}
\end{equation}
(the perturbed Stein-Stein model),
\begin{equation}
\left\{\begin{array}{ll}
d{\widetilde X}_t=\mu {\widetilde X}_{t-}dt+\sqrt{Y_t}{\widetilde X}_{t-}dW_t+{\widetilde X}_{t-}dJ_t\\
dY_t=q\left(m-Y_t\right)dt+c\sqrt{Y_t}dZ_t,
\end{array}
\right.
\label{E:H1}
\end{equation}
(the perturbed Heston model), and 
\begin{equation}
\left\{\begin{array}{l}
d{\widetilde X}_t=\mu {\widetilde X}_{t-}dt+Y_t{\widetilde X}_{t-} dW_t+{\widetilde X}_{t-}dJ_t\\
dY_t=\nu Y_tdt+\xi Y_tdZ_t,
\end{array}
\right.
\label{E:sde2}
\end{equation}
(the perturbed Hull-White model). It is assumed in (\ref{E:SS1}), (\ref{E:H1}), and (\ref{E:sde2}) that the compound Poisson process 
$J$ is independent of standard Brownian motions $W$ and $Z$.

We will next formulate several results of Gulisashvili and Stein. For the uncorrelated Heston model, there exist constants 
$A_1> 0$, $A_2> 0$, and $A_3> 2$ such that

\begin{equation}
D_t(x)=A_1(\log x)^{-\frac{3}{4}+\frac{qm}{c^2}}e^{A_2\sqrt{\log x}}x^{-A_3}
\left(1+O\left((\log x)^{-\frac{1}{4}}\right)\right)
\label{E:dopos}
\end{equation}
as $x\rightarrow\infty$ (see \cite{GS09-3}). For the uncorrelated Stein-Stein model, there exist constants $B_1> 0$, $B_2> 0$, 
and $B_3> 2$ such that

\begin{equation}
D_t(x)=B_1(\log x)^{-\frac{1}{2}}e^{B_2\sqrt{\log x}}x^{-B_3}
\left(1+O\left((\log x)^{-\frac{1}{4}}\right)\right)
\label{E:dop1s}
\end{equation}
as $x\rightarrow\infty$ (see \cite{GS09-3}). Finally, in the case of the uncorrelated Hull-White model, 
there exist constants $b_1> 0$, $b_2$ and $b_3$ such that following formula holds (see \cite{GS09-2} and also Theorem 4.1 
in \cite{GS09-3}):

\begin{align}
&D_t(x)=
b_1x^{-2}(\log x)^{\frac{b_2-1}{2}}
\left(\log\log x\right)^{b_3} \nonumber \\
&\quad\exp\left\{-\frac{1}{2t\xi^2}\left(\log\left[\frac{1}{y_0}\sqrt{\frac{2\log x}{t}}\right]
+\frac{1}{2}\log\log\left[\frac{1}{y_0}\sqrt{\frac{2\log x}{t}}\right]\right)^2\right\} \nonumber \\
&\quad\left(1+O\left((\log\log x)^{-\frac{1}{2}}\right)\right)
\label{E:main3s}
\end{align}
as $x\rightarrow\infty.$ The constants in formulas (\ref{E:dopos}), (\ref{E:dop1s}), and (\ref{E:main3s}) depend on the model parameters.  Explicit expressions for these 
constants can be found in \cite{GS09-2} and \cite{GS09-3}. The constants $A_3$ and $B_3$, appearing in 
(\ref{E:dopos}) and (\ref{E:dop1s}), 
describe the rate of the power-type decay of the stock price distribution density in the Heston and the Stein-Stein model, respectively. 
The explicit formulas for these constants are as follows: 

\begin{equation}
A_3=\frac{3}{2}+\frac{\sqrt{8C+t}}{2\sqrt{t}}\quad
\mbox{with}\quad
C=\frac{t}{2c^2}\left(q^2+\frac{4}{t^2}r^2_{\frac{qt}{2}}\right),
\label{cons1}
\end{equation}
and

\begin{equation}
B_3=\frac{3}{2}+\frac{\sqrt{8G+t}}{2\sqrt{t}}\quad
\mbox{with}\quad
G=\frac{t}{2\sigma^2}\left(q^2+\frac{1}{t^2}r^2_{qt}\right).
\label{cons2}
\end{equation}
In (\ref{cons1}) and (\ref{cons2}), 
$r_s$ denotes the smallest positive root of the entire function 
$$
z\mapsto z\,cos z+s\,sin z.
$$ 
Formulas (\ref{cons1}) and (\ref{cons2}) can be found in \cite{GS09-3}.

The distribution density density $D_t$ in uncorrelated stochastic volatility models satisfies the following symmetry condition:
\begin{equation}
\left(\frac{x_0e^{\mu t}}{x}\right)^3D_t\left(\frac{\left(x_0e^{\mu t}\right)^2}{x}\right)=D_t(x),\quad x> 0,
\label{E:sym}
\end{equation}
(see Section 2 in \cite{GS09-3}). This condition shows that the asymptotic behavior of the stock price distribution density near zero is completely
determined by its behavior near infinity.
\section{Main results}\label{MR}

The following theorems concern the tail behavior of the stock price distribution 
density in perturbed Stein-Stein, Heston, and Hull-White models:
\begin{theorem}\label{T1}

Let $\varepsilon>0.$ Then there exist $c_1>0$, $c_2>0,$ and $x_1>0$ such that the following estimates hold for the distribution density 
${\widetilde D}_t$ of the stock price $\widetilde{X}_t$ in the perturbed Heston model:

\begin{equation}
c_1\left(\frac{1}{x^{A_3}}+\frac{1}{x^{1+\eta_1}}\right)\leq {\widetilde D}_t (x)
\leq c_2\left(\frac{1}{x^{A_3-\varepsilon}}+\frac{1}{x^{1+\eta_1-\varepsilon}}\right)
\label{mf1}
\end{equation}
for all $x>x_1.$ In (\ref{mf1}), the constant $A_3$ is given by (\ref{cons1})
and the constants $c_2$ and $x_1$ depend on $\varepsilon.$ 

\end{theorem}

\begin{theorem}\label{T2}

Let $\varepsilon>0.$ Then there exist $c_3>0$, $c_4>0,$ and $x_2>0$ such that the following estimates hold for the distribution density 
${\widetilde D}_t$ of the stock price $\widetilde{X}_t$ in the perturbed Heston model:

\begin{equation}
c_3\left(x^{A_3-3}+x^{\eta_2-1}\right)\leq {\widetilde D}_t (x)
\leq c_4\left(x^{A_3-3-\varepsilon}+x^{\eta_2-1-\varepsilon}\right)
\label{mf2}
\end{equation}
for all $0<x<x_2.$ Here the constant $A_3$ is the same as in Theorem \ref{T1}
and the constants $c_4$ and $x_2$ depend on $\varepsilon.$ 

\end{theorem}

\begin{theorem}\label{T3}

Let $\varepsilon>0.$ Then there exist $c_5>0$, $c_6>0,$ and $x_3>0$ such that the following estimates hold for 
the distribution density ${\widetilde D}_t$ of the stock price $\widetilde{X}_t$ in the perturbed Stein-Stein model:

\begin{equation}
c_5\left(\frac{1}{x^{B_3}}+\frac{1}{x^{1+\eta_1}}\right)\leq {\widetilde D}_t (x)
\leq c_6\left(\frac{1}{x^{B_3-\varepsilon}}+\frac{1}{x^{1+\eta_1-\varepsilon}}\right)
\label{mf3}
\end{equation}
for all $x>x_3.$ In (\ref{mf3}), the constant $B_3$ is given by (\ref{cons2}) and
the constants $c_6$ and $x_3$ depend on $\varepsilon.$ 

\end{theorem}

\begin{theorem}\label{T4}

Let $\varepsilon>0.$ Then there exist $c_7>0$, $c_8>0,$ and $x_4>0$ such that the following estimates hold for 
the distribution density ${\widetilde D}_t$ of the stock price $\widetilde{X}_t$ in the perturbed Stein-Stein model:

\begin{equation}
c_7\left(x^{B_3-3}+x^{\eta_2-1}\right)\leq {\widetilde D}_t (x)
\leq c_8\left(x^{B_3-3-\varepsilon}+x^{\eta_2-1-\varepsilon}\right)
\label{mf4}
\end{equation}
for all $0<x<x_4.$ Here the constant $B_3$ is the same as in Theorem \ref{T3} and
the constants $c_8$ and $x_4$ depend on $\varepsilon.$ 

\end{theorem}

We will prove Theorems \ref{T1}-\ref{T4} in Section \ref{P}. In the remaining part of the present section, we compare the tail 
behavior of the stock price distribution density before and after perturbation by a compound Poisson process.


Let us begin with the Heston model. It follows from Theorem \ref{T1} that if $1+\eta_1<A_3$, then 

$$
\frac{{\bar c}_1}{x^{1+\eta_1}}\leq {\widetilde D}_t (x)
\leq \frac{{\bar c}_2}{x^{1+\eta_1-\varepsilon}},\quad x>{\bar x}_1.
$$
Therefore, formula (\ref{E:dopos}) shows that that if the condition $1+\eta_1<A_3$ holds, then the tail of the distribution of the stock price in the perturbed Heston model is heavier than in the original model. 

On the other hand, if 
$1+\eta_1>A_3$, then Theorem \ref{T1} implies the following estimate:  

$$
\frac{{\tilde c}_1}{x^{A_3}}\leq {\widetilde D}_t(x)
\leq \frac{{\tilde c}_2}{x^{A_3-\varepsilon}},\quad x>{\tilde x}_1.
$$
Now formula (\ref{E:dopos}) shows that if $1+\eta_1>A_3$, then there are no significant changes 
in the tail behavior of the distribution density of the stock price after perturbation. Similar assertions hold for the Stein-Stein model. This can be established 
using Theorem \ref{T3} and formula (\ref{E:dop1s}). 

Next, suppose $x\rightarrow 0$. Then we can compare the behavior of the distribution density of the stock price 
in unperturbed and perturbed models, taking into account Theorem \ref{T2}, Theorem \ref{T4}, formula (\ref{E:dopos}), 
formula (\ref{E:dop1s}), and the symmetry condition (\ref{E:sym}).
For instance, if $\eta_2<A_3-2$ in the perturbed Heston model, then 

$$
{\bar c}_3 x^{\eta_2-1}\leq {\widetilde D}_t (x)
\leq {\bar c}_4 x^{\eta_2-1-\varepsilon}
$$ 
for all $x<{\bar x}_2.$ On the other hand if $\eta_2>A_3-2$, then 

$$
{\tilde c}_3 x^{A_3-3}\leq {\widetilde D}_t (x)
\leq {\tilde c}_4 x^{A_3-3-\varepsilon}
$$
for all $x<{\tilde x}_2.$ Similar results hold for the Stein-Stein model. 

For the Hull-White model, there are no significant changes in the tail behavior of the stock price distribution after perturbation.
This statement follows from the assumption 
$\eta_1> 1$ and from the fact that the stock price density in the unperturbed Hull-White model 
decays like $x^{-2}$ (see formula (\ref{E:main3s})). 
\section{Proofs of the main results}\label{P}
The proofs of Theorems \ref{T1}-\ref{T4} are based on an explicit formula for the distribution density $\widetilde{D}_t$ of the 
stock price $\widetilde{X}_t$ in perturbed Heston, Stein-Stein, and Hull-White models (see formula (\ref{density}) below). 
Note that the stock price process $\widetilde{X}$ 
in the perturbed Stein-Stein and Hull-White models is given by 

\begin{equation}
{\widetilde X}_t=x_0\exp\left\{\mu t-\frac{1}{2}\int_0^t Y_s^2 ds+\int_0^t Y_s dW_s+\sum_{i=1}^{N_t} U_i \right\},
\label{D1}
\end{equation}
while for the perturbed for Heston model we have

\begin{equation}
{\widetilde X}_t=x_0\exp\left\{\mu t-\frac{1}{2}\int_0^t Y_s ds+\int_0^t \sqrt{Y_s} dW_s+\sum_{i=1}^{N_t} U_i \right\}.
\label{D2}
\end{equation}
Formulas (\ref{D1}) and (\ref{D2}) can be established using the Doléans-Dade formula (see, for example, \cite{P05}). 
We will denote by $\mu_t$ the distribution of the random variable $J_t$ defined in (\ref{Poi}). 
It is not hard to see that the following formula holds:

\begin{equation}
\mu_t (A)=\pi_0 \delta_0 (A)+\sum_{n=1}^{\infty} \pi_n \int_A f^{*(n)}(u)du
\label{jm}
\end{equation}
where 
$\pi_0=e^{-\lambda t}$, $\pi_n=e^{-\lambda t}(n!)^{-1} (\lambda t)^n$ for $n\geq 1$, 
$A$ is a Borel subset of $\mathbb R$, and $f$ is given by (\ref{DE}). The star in (\ref{jm}) denotes the convolution. 

The distribution density $D_t$ of the stock price $X_t$ in uncorrelated models of our interest is related to the 
law of the following random variable: 
$$
\alpha_t=\left\{\frac{1}{t}\int_0^t Y_s^2 ds\right \}^{\frac{1}{2}}
$$
for the Stein-Stein and the Hull-White model, and 
$$
\alpha_t=\left\{\frac{1}{t}\int_0^t Y_s ds\right \}^{\frac{1}{2}}
$$
for the Heston model (see \cite{GS09-2} and \cite{GS09-3}). The distribution density of the random variable $\alpha_t$ is 
called the mixing distribution density 
and is denoted by $m_t$. 
We refer the reader to \cite{GS09-2}, \cite{GS09-3}, and \cite{SS91} for more information on the mixing distribution density. 

The next lemma establishes a relation between the mixing distribution density $m_t$ in the uncorrelated model and 
the distribution density ${\widetilde D}_t$ of the stock price ${\widetilde X}_t$ in the corresponding perturbed model.
\begin{lemma}\label{L1}
The density ${\widetilde D}_t$ in perturbed Stein-Stein, Heston and Hull-White models is given by the following formula:

$$
{\widetilde D}_t (x)=\frac{1}{\sqrt{2\pi t} x} \int_0^{\infty}\left (\int_{\mathbb R} \exp\left
\{-\frac{(\log \frac{x}{x_0 e^{\mu t}}+\frac{ty^2}{2}-u)^2}{2ty^2}\right\}\mu_t (du)\right)m_t(y)\frac{dy}{y},
$$
where $m_t$ is the mixing distribution density and $\mu_t$ is defined by (\ref{jm}).
\end{lemma}

\begin{proof}
We will prove Lemma \ref{L1} for the Heston model. The proof for the Stein-Stein and the Hull-White model is similar. 
For the latter models, we use formula (\ref{D1}) instead of formula (\ref{D2}). 

Put $T_t=\sum_{i=1}^{N_t} U_i$. Then for any $\eta>0$, formula (\ref{D2}) gives 

\begin{eqnarray*}
{\mathbb P}\left(\widetilde{X}_t\leq \eta\right)&=&{\mathbb P}\left[\int_0^t \sqrt{Y_s} dW_s
+T_t\leq \log{\frac{\eta}{x_0 e^{\mu t}}}+\frac{t\alpha_t^2}{2}\right]\\
&=&{\mathbb E}\int_{-\infty}^{z_*}\int_{-\infty}^{\infty} \frac{1}{\sqrt{2\pi t} \alpha_t}
\exp\left\{-\frac{(z-u)^2}{2t \alpha_t^2}\right\}\mu_t (du) dz,
\end{eqnarray*}
where $\displaystyle{z_*=\log\frac{\eta}{x_0 e^{\mu t}}+\frac{t \alpha_t^2}{2}}$.
Making the substitution 
$\displaystyle{z=\log\frac{x}{x_0 e^{\mu t}}+\frac{t \alpha_t^2}{2}}$,
we obtain

\begin{align*}
&{\mathbb P}\left(\widetilde{X}_t\leq \eta\right) \\
&={\mathbb E}\int_{-\infty}^{\eta}\int_{-\infty}^{\infty} \frac{1}{\sqrt{2\pi t} \alpha_t}
\exp\left\{-\frac{(\log \frac{x}{x_0 e^{\mu t}}+\frac{t\alpha_t^2}{2}-u)^2}{2t \alpha_t^2}\right\}\mu_t (du) \frac{dx}{x} \\
&=\int_{-\infty}^{\eta}\int_0^{\infty}\int_{-\infty}^{\infty}
\exp\left\{-\frac{(\log \frac{x}{x_0 e^{\mu t}}+\frac{ty^2}{2}-u)^2}{2ty^2}\right\}\mu_t (du) 
\frac{m_t(y)}{\sqrt{2\pi t} y}dy\frac{dx}{x}.
\end{align*}

It is clear that the previous equality implies Lemma \ref{L1}.
\end{proof}
\begin{remark} \rm
It follows from Lemma \ref{L1} that 

\begin{align}\label{density}
&\widetilde{D}_t(x)=\frac{\sqrt{x_0 e^{\mu t}}}{\sqrt{2\pi t} x^{\frac{3}{2}}} \nonumber \\ 
&\int_{\mathbb R}e^{\frac{u}{2}}\mu_t (du) \int_0^{\infty}
\frac{m_t (y)}{y} \exp\left\{-\frac{(\log \frac{x}{x_0 e^{\mu t}}-u)^2}{2ty^2}-\frac{ty^2}{8}\right\}dy .
\end{align}
This representation will be used below to obtain two-sided estimates for the distribution density of the stock price in 
perturbed stochastic volatility models.
\end{remark}

\bf Proof of Theorem \ref{T1}. \rm 
The next lemma will be needed in the proof of Theorem \ref{T1}. 
\begin{lemma}\label{L:kou}
Let $f$ be the density of the double exponential law (see formula (\ref{DE})). Then for every $n> 1$, the following formula holds:
\begin{align}
f^{*(n)}(u)&=e^{-\eta_1u}\sum_{k=1}^nP_{n,k}\eta_1^k\frac{1}{(k-1)!}u^{k-1}{1\!\!1}_{\{u\ge 0\}}\nonumber\\
&\quad+e^{\eta_2u}\sum_{k=1}^nQ_{n,k}\eta_2^k\frac{1}{(k-1)!}(-u)^{k-1}{1\!\!1}_{\{u< 0\}},
\label{E:k1}
\end{align}
where
$$
P_{n,k}=\sum_{i=k}^{n-1}\binom{n-k-1}{i-k}\binom{n}{i}\left(\frac{\eta_1}{\eta_1+\eta_2}\right)^{i-k}
\left(\frac{\eta_2}{\eta_1+\eta_2}\right)^{n-i}p^i q^{n-i}
$$
for all $1\leq k\leq n-1,$ and 

$$
Q_{n,k}=\sum_{i=k}^{n-1}\binom{n-k-1}{i-k}\binom{n}{i}\left(\frac{\eta_1}{\eta_1+\eta_2}\right)^{n-i}
\left(\frac{\eta_2}{\eta_1+\eta_2}\right)^{i-k}p^{n-i} q^i
$$ for all $1\leq k\leq n-1.$
In addition, $P_{n,n}=p^n$ and $Q_{n,n}=q^n.$ 
\end{lemma}

Lemma \ref{L:kou} can be established using Proposition B.1 in \cite{K02} and taking into account simple properties of the exponential distribution.  

The next statement follows from Lemma \ref{L:kou} and formula (\ref{jm}):

\begin{lemma}\label{jm2}
For every Borel set $A\subset {\mathbb R},$ 

\begin{equation}
\mu_t(A)=\pi_0 \delta_0(A)+\int_{A\cap [0,\infty)} G_1(u)e^{-\eta_1 u}du+\int_{A\cap (-\infty, 0)} G_2(u)e^{\eta_2 u}du,
\label{jmf1}
\end{equation}
where 

\begin{equation}
G_1(u)=\sum_{k=0}^{\infty} \left[\frac{\eta_1^{k+1}}{k!} \sum_{n=k+1}^{\infty} \pi_n P_{n,k+1}\right]u^k,
\label{jmf2}
\end{equation}
and 

\begin{equation}
G_2(u)=\sum_{k=0}^{\infty} \left[\frac{\eta_2^{k+1}}{k!} \sum_{n=k+1}^{\infty} \pi_n Q_{n,k+1}\right](-u)^k.
\label{jmf3}
\end{equation}

\end{lemma}

Our next goal is to estimate the rate of growth of the functions $G_1$ and $G_2$ defined by (\ref{jmf2}) and (\ref{jmf3}).

\begin{lemma}\label{L:gr}
For every $\varepsilon>0$ the function $G_1$ grows slower than the function $u\mapsto e^{\varepsilon u}$ as $u\rightarrow \infty.$ Similarly, the function $G_2$ grows slower than the function $u\mapsto e^{-\varepsilon u}$ as $u\rightarrow -\infty.$
\end{lemma}

\begin{proof}
We will prove the lemma by comparing the Taylor coefficients 

$$a_k=\frac{1}{k!}\eta_1^{k+1} \sum_{n=k+1}^{\infty} \pi_n P_{n,k+1},\quad k\ge 0,$$
of the function $G_1$ and the Taylor coefficients $b_k=\frac{1}{k!}\varepsilon^k$, $k\ge 0$,
of the function $e^{\varepsilon u}.$ We have
$a_k\leq b_k$ for $k>k_0$.
The previous inequality can be established using the estimate
$$
\eta_1^{k+1} \sum_{n=k+1}^{\infty} \pi_n P_{n,k+1}\leq \eta_1^{k+1} \sum_{n=k+1}^{\infty} \pi_n,
$$
and taking into account the fast decay of the complementary distribution function of the Poisson distribution.

This completes the proof of Lemma \ref{L:gr} for the function $G_1$. 
The proof for the function $G_2$ is similar. 

\end{proof}

The following lemma was obtained in \cite{GS09-3} (formula (54)): 
\begin{lemma}\label{L10}  
Let $m_t$ be the mixing distribution density in the Heston model. Then there exist constants $H_1> 0$ and $H_2> 0,$ depending on the model parameters,  such that 
\begin{align*}
&\int_0^{\infty} \frac{m_t (y)}{y}\exp\left\{-\left(\frac{\omega^2}{2ty^2}+\frac{ty^2}{8}\right)\right\}dy\\
&=H_1 \omega^{-\frac{3}{4}+\frac{qm}{c^2}} e^{H_2 \sqrt{\omega}}\exp\left\{-\frac{\sqrt{8C+t}}{2\sqrt{t}}\omega\right\}
\left(1+O\left(\omega^{-\frac{1}{4}}\right)\right)
\end{align*}
as $\omega\rightarrow\infty$. The constant $C$ in the previous formula is given by (\ref{cons1}).
\end{lemma}

{\bf Proof of the estimate from below in Theorem \ref{T1}}. \rm We will use formula (\ref{density}) in the proof. 
Put $z=\log \frac{x}{x_0 e^{\mu t}}.$ Then we have

\begin{equation}\label{den2}
\widetilde{D}_t(x)=\frac{\sqrt{x_0 e^{\mu t}}}{\sqrt{2\pi t} x^{\frac{3}{2}}} 
\int_{\mathbb R}e^{\frac{u}{2}}\mu_t (du) \int_0^{\infty}
\frac{m_t (y)}{y} \exp\left\{-\frac{(z-u)^2}{2ty^2}-\frac{ty^2}{8}\right\}dy.
\end{equation}
Note that for the uncorrelated Heston model the following formula holds: 
\begin{equation}\label{den3}
D_t(x)=\frac{\sqrt{x_0 e^{\mu t}}}{\sqrt{2\pi t} x^{\frac{3}{2}}}\int_0^{\infty}\frac{m_t (y)}{y} 
\exp\left\{-\frac{z^2}{2ty^2}-\frac{ty^2}{8}\right\}dy
\end{equation}(see \cite{GS09-3}).

Let $\rho$ be any increasing function of $z$ such that $\rho(z)< z$ and $z-\rho(z)\rightarrow \infty$
as $z\rightarrow \infty$. Then (\ref{den2}) gives 

\begin{equation}\label{den4}
\widetilde{D}_t(x)\geq I_1+I_2,
\end{equation}
where
\begin{equation}
I_1=\frac{\sqrt{x_0 e^{\mu t}}}{\sqrt{2\pi t} x^{\frac{3}{2}}}\int_1^{\rho(z)} e^{\frac{u}{2}}\mu_t (du) \int_0^{\infty}
\frac{m_t (y)}{y} \exp\left\{-\frac{(z-u)^2}{2ty^2}-\frac{ty^2}{8}\right\}dy
\label{den5}
\end{equation}
and
\begin{equation}
I_2=\frac{\sqrt{x_0 e^{\mu t}}}{\sqrt{2\pi t} x^{\frac{3}{2}}}\int_z^{z+1}e^{\frac{u}{2}}\mu_t (du) \int_0^{\infty}
\frac{m_t (y)}{y} \exp\left\{-\frac{(z-u)^2}{2ty^2}-\frac{ty^2}{8}\right\}dy.
\label{den6}
\end{equation}

Throughout the remaining part of the section, we will denote by $\alpha$ a positive constant which may differ from line to line. 
Since the function $G_1$ is increasing on $(0,\infty)$ and (\ref{jmf1}) and (\ref{jmf2}) hold, we have 
$$
I_2\ge \alpha x^{-\frac{3}{2}}\int_z^{z+1}e^{\frac{u}{2}}e^{-\eta_1 u}du\int_0^{\infty}
\frac{m_t (y)}{y} \exp\left\{-\frac{1}{2ty^2}-\frac{ty^2}{8}\right\}dy,\quad x>x_0.
$$
It is known that $\displaystyle{\int_0^1y^{-1}m_t(y)dy<\infty}$ (see \cite{GS09-3}). Therefore, the second integral in the previous estimate converges. 
It follows that
$$
I_2\ge \alpha x^{-\frac{3}{2}}\int_z^{z+1}e^{\frac{u}{2}}e^{-\eta_1 u}du=cx^{-1-\eta_1}
$$
for $x> x_0$. It is not hard to see using the inequality ${\tilde D}_t(x)\geq I_2$ that the estimate from below in (\ref{mf1}) holds 
in the case where $1+\eta_1\leq A_3.$

It remains to prove the estimate from below under the assumption $1+\eta_1> A_3.$ We will use the inequality 
${\tilde D}_t(x)\geq I_1$ in the proof. To estimate $I_1$ we notice that $z-u\geq z-\rho(z)\rightarrow \infty$ as $x\rightarrow \infty.$ 
Therefore, Lemma \ref{L10} can be applied to estimate the second integral on the right-hand side of (\ref{den5}). This gives 
\begin{align*}
I_1&\geq \alpha x^{-\frac{3}{2}}\int_1^{\rho(z)} e^{\frac{u}{2}}G_1 (u)e^{-\eta_1 u}
(z-u)^{-\frac{3}{4}+\frac{qm}{c^2}} \\
&\quad e^{H_2 \sqrt{z-u}}
\exp\left\{-\frac{\sqrt{8C+t}}{2\sqrt{t}}(z-u)\right\}du.
\end{align*}
Since the function $G_1$ is increasing on $(0,\infty)$ and the function 
$$
y\mapsto y^{-\frac{3}{4}+\frac{qm}{c^2}} e^{H_2 \sqrt{y}}
$$ 
is eventually increasing, the previous inequality gives
\begin{align*}
&I_1\geq \alpha x^{-\frac{3}{2}}\int_1^{\rho(z)} e^{\frac{u}{2}}e^{-\eta_1 u}
\exp\left\{-\frac{\sqrt{8C+t}}{2\sqrt{t}}(z-u)\right\}du\\
&=\alpha x^{-A_3}\int_1^{\rho(z)} \exp\left\{\left(A_3-1-\eta_1\right)u\right\}du.
\end{align*}
Here we used the equality $A_3=\frac{3}{2}+\frac{\sqrt{8C+t}}{2\sqrt{t}}$ (see (\ref{cons1})).
Since $A_3<1+\eta_1$ and $\rho(z)\rightarrow \infty$ as $z\rightarrow \infty$, we get $I_1\geq \alpha x^{-A_3},\quad x>x_0.$
This establishes the estimate from below in Theorem \ref{T1} in the case where $A_3<1+\eta_1$. 
\\

{\bf Proof of the estimate from above in Theorem \ref{T1}}. \rm Let $\varepsilon$ be a small positive number. Denote by $\Lambda_t(z,u)$ the following integral:
$$
\int_0^{\infty}
\frac{m_t (y)}{y} \exp\left\{-\frac{(z-u)^2}{2ty^2}-\frac{ty^2}{8}\right\}dy,
$$
Then formula 
(\ref{den2}) can be rewritten as follows:
\begin{equation}
\widetilde{D}_t(x)=\frac{\sqrt{x_0 e^{\mu t}}}{\sqrt{2\pi t} x^{\frac{3}{2}}} 
\int_{\mathbb R}e^{\frac{u}{2}}\Lambda_t(z,u)\mu_t (du)=J_1+J_2+J_3,
\label{E:nu}
\end{equation}
where 
$$
J_1=\frac{\sqrt{x_0 e^{\mu t}}}{\sqrt{2\pi t} x^{\frac{3}{2}}} 
\int_{-\infty}^{0-}e^{\frac{u}{2}}\Lambda_t(z,u)\mu_t (du),
$$
$$
J_2=\frac{\sqrt{x_0 e^{\mu t}}}{\sqrt{2\pi t} x^{\frac{3}{2}}} 
\int_0^{sz}e^{\frac{u}{2}}\Lambda_t(z,u)\mu_t (du),
$$
and
$$
J_3=\frac{\sqrt{x_0 e^{\mu t}}}{\sqrt{2\pi t} x^{\frac{3}{2}}} 
\int_{sz}^{\infty}e^{\frac{u}{2}}\Lambda_t(z,u)\mu_t (du).
$$
The number $s$ in the previous equalities satisfies $0< s< 1$. The value of $s$ will be chosen below.

To estimate $J_2$, we notice that if $x$ is large, then $z-u$ in the expression for $J_2$ is also large. Using Lemma \ref{jm2} 
and Lemma \ref{L10}, we see that
\begin{align*}
&J_2\leq \alpha D_t(x)+\alpha x^{-\frac{3}{2}}\int_0^{sz} e^{\frac{u}{2}}G_1 (u)e^{-\eta_1 u}
(z-u)^{-\frac{3}{4}+\frac{qm}{c^2}} e^{H_2 \sqrt{z-u}}\\
&\quad\exp\left\{-\frac{\sqrt{8C+t}}{2\sqrt{t}}(z-u)\right\}du.
\end{align*}
Since the functions $G_1(y)$ and $y\mapsto y^{-\frac{3}{4}+\frac{qm}{c^2}} e^{H_2 \sqrt{y}}$ grow slower than the function $y\mapsto\exp\left\{\frac{\varepsilon}{2}y\right\}$ (see Lemma \ref{L:gr}), the previous inequality and formula 
(\ref{E:dopos}) imply that
\begin{align}
&J_2\leq\alpha x^{-A_3+\varepsilon}+\alpha x^{-\frac{3}{2}}\int_0^{sz}\exp\left\{\left(\frac{1}{2}-\eta_1
+\frac{\varepsilon}{2}\right)u\right\} \nonumber \\
&\quad\exp\left\{\left(-\frac{\sqrt{8C+t}}{2\sqrt{t}}+\frac{\varepsilon}{2}\right)(z-u)\right\}du \nonumber \\
&\le\alpha x^{-A_3+\varepsilon}+\alpha x^{-A_3+\frac{\varepsilon}{2}}\int_0^z
\exp\left\{\left(A_3-1-\eta_1\right)u\right\}du \nonumber \\
&\le\alpha\left(\frac{1}{x^{A_3-\varepsilon}}+\frac{1}{x^{1+\eta_1-\varepsilon}}\right)
\label{E:ue1}
\end{align}
for $x> x_0$.

The function $\Lambda_t$ is bounded (this has already been established in the previous part of the proof). Therefore,
\begin{equation}
J_3\leq\alpha x^{-\frac{3}{2}}\int_{sz}^{\infty}e^{\frac{u}{2}}G_1(u)e^{-\eta_1 u}du.
\label{E:ue2}
\end{equation}
Since the function $G_1(u)$ grows slower than the function $y\mapsto\exp\left\{\zeta u\right\}$ for any $\zeta> 0$ 
(see Lemma \ref{L:gr}), estimate (\ref{E:ue2}) implies that
$$
J_3\leq\alpha x^{-\frac{3}{2}+s\left(\frac{1}{2}+\zeta-\eta_1\right)},\quad x> x_0.
$$
Now using the fact that $\zeta$ can be any close to 0 and s any close to 1, we see that
\begin{equation}
J_3\leq\alpha\frac{1}{x^{1+\eta_1-\varepsilon}},\quad x> x_0.
\label{E:ue3}
\end{equation}

We will next estimate $J_1$. It follows from Lemma \ref{jm2} that
$$
J_1=\alpha x^{-\frac{3}{2}}\int_{-\infty}^{0-}e^{\frac{u}{2}}\Lambda_t(z,u)G_2(u)e^{\eta_2 u}du.
$$
Since $u< 0$, we see that $z-u$ is large if $x$ is large. Using Lemma \ref{L10}, we obtain
\begin{align}
J_1&\leq\alpha x^{-\frac{3}{2}}\int_{-\infty}^{0-}e^{\frac{u}{2}}(z-u)^{-\frac{3}{4}+\frac{qm}{c^2}} e^{H_2 \sqrt{z-u}}
\nonumber \\
&\quad\exp\left\{-\frac{\sqrt{8C+t}}{2\sqrt{t}}(z-u)\right\}G_2(u)e^{\eta_2 u}du.
\label{E:ue4}
\end{align}
The function $y\mapsto y^{-\frac{3}{4}+\frac{qm}{c^2}} e^{H_2 \sqrt{y}}$ is eventually increasing. Moreover, 
it grows slower than $e^{\frac{\epsilon}{2}y}$. Since $z-u> z$ in (\ref{E:ue4}), we have
\begin{align}
&J_1\leq\alpha x^{-\frac{3}{2}}\int_{-\infty}^{0-}e^{\frac{u}{2}}
\exp\left\{\left(-\frac{\sqrt{8C+t}}{2\sqrt{t}}+\frac{\varepsilon}{2}\right)(z-u)\right\}G_2(u)e^{\eta_2 u}du
\nonumber \\
&\leq \alpha x^{-A_3+\frac{\varepsilon}{2}}\int_{-\infty}^{0-}e^{\frac{u}{2}}
\exp\left\{\left(\frac{\sqrt{8C+t}}{2\sqrt{t}}-\frac{\varepsilon}{2}\right)u\right\}G_2(u)e^{\eta_2 u}du \nonumber \\
&=\alpha x^{-A_3+\frac{\varepsilon}{2}}\int_0^{\infty}
\exp\left\{\left(-\frac{1}{2}-\eta_2-\frac{\sqrt{8C+t}}{2\sqrt{t}}+\frac{\varepsilon}{2}\right)u\right\}G_2(-u)du. 
\label{E:ue5}
\end{align}
If $\varepsilon$ is sufficiently small, then the integral in (\ref{E:ue5}) converges (use Lemma \ref{L:gr}). It follows from (\ref{E:ue5}) that
\begin{equation}
J_1\leq \alpha\frac{1}{x^{A_3-\varepsilon}},\quad x> x_0.
\label{E:ue6}
\end{equation}

Finally, combining (\ref{E:nu}), (\ref{E:ue1}), (\ref{E:ue3}), and (\ref{E:ue6}), we establish the estimate from above in Theorem \ref{T1}.
\\
\\
\bf Proof of Theorem \ref{T2}. \rm The following formula can be obtained from (\ref{density}):
\begin{align}\label{densite}
&\left(\frac{x_0e^{\mu t}}{x}\right)^3\widetilde{D}_t\left(\frac{\left(x_0e^{\mu t}\right)^2}{x}\right)
=\frac{\sqrt{x_0 e^{\mu t}}}{\sqrt{2\pi t} x^{\frac{3}{2}}} \nonumber \\ 
&\int_{\mathbb R}e^{\frac{u}{2}}\mu_t (du) \int_0^{\infty}
\frac{m_t (y)}{y} \exp\left\{-\frac{(\log \frac{x}{x_0 e^{\mu t}}+u)^2}{2ty^2}-\frac{ty^2}{8}\right\}dy .
\end{align}
It follows from (\ref{densite}) and (\ref{jmf1}) that
\begin{align}\label{densito}
&\left(\frac{x_0e^{\mu t}}{x}\right)^3\widetilde{D}_t\left(\frac{\left(x_0e^{\mu t}\right)^2}{x}\right)
=\frac{\sqrt{x_0 e^{\mu t}}}{\sqrt{2\pi t} x^{\frac{3}{2}}} \nonumber \\ 
&\int_{\mathbb R}e^{\frac{u}{2}}\tilde{\mu}_t (du) \int_0^{\infty}
\frac{m_t (y)}{y} \exp\left\{-\frac{(\log \frac{x}{x_0 e^{\mu t}}-u)^2}{2ty^2}-\frac{ty^2}{8}\right\}dy,
\end{align}
where 
\begin{align}
&\tilde{\mu}_t(A)=\pi_0\delta_0(A)+\int_{A\cap(0,\infty)}G_2(-u)e^{-\left(\eta_2+1\right)u}du \nonumber \\
&\quad+\int_{A\cap(-\infty,0)}G_1(-u)e^{\left(\eta_1-1\right)u}du
\label{E:de1}
\end{align}
for all Borel sets $A\subset\mathbb{R}$. In (\ref{E:de1}), $G_1$ and $G_2$ are defined by (\ref{jmf2}) and
(\ref{jmf3}), respectively. Now it is clear that we can use the proof of Theorem \ref{T1} with the 
pairs $\left(\eta_1,p\right)$ and $\left(\eta_2,q\right)$ replaced by the pairs $\left(\eta_2+1,q\right)$
and $\left(\eta_1-1,p\right)$, respectively. We should also take into account Lemma \ref{L:gr}. 
It is not hard to 
see using (\ref{densite}) that for every $\varepsilon> 0$, there exist constants $\tilde{c}_1> 0$, $\tilde{c}_2> 0$,
and $\tilde{x}> 0$ such that the following estimates hold:
\begin{equation}
\tilde{c}_1\left(\frac{1}{x^{A_3}}+\frac{1}{x^{\eta_2+2}}\right)\le x^{-3}\widetilde{D}_t\left(\frac{\left(x_0e^{\mu t}\right)
^2}{x}\right)\le\tilde{c}_2\left(\frac{1}{x^{A_3-\varepsilon}}+\frac{1}{x^{\eta_2+2-\varepsilon}}\right)
\label{E:de2}
\end{equation}
for all $x>\tilde{x}$. The constants $\tilde{c}_2$ and $\tilde{x}$ depend on $\varepsilon$. Now it is clear that
(\ref{mf2}) follows from (\ref{E:de2}).

This completes the proof of Theorem \ref{T2}

We do not include the proofs of Theorems \ref{T3} and \ref{T4}, because these theorems can be established exactly as Theorems \ref{T1}
and \ref{T2}.

\end{document}